\pdfoutput=1
\documentclass[conference]{IEEEtran}
\usepackage[utf8]{inputenc} 
\usepackage[T1]{fontenc} 
\usepackage[english]{babel} 
\usepackage{tabularx,booktabs}
\newcolumntype{C}{>{\centering\arraybackslash}X} 
\usepackage{csquotes}
\MakeOuterQuote{"}

\usepackage{amsmath}
\usepackage{amsfonts}
\usepackage{amssymb}
\usepackage{graphicx}
\usepackage{subcaption}
\usepackage{textcomp}
\usepackage{float}
\usepackage{setspace}
\usepackage{url}

\usepackage[final]{pdfpages}
\usepackage{multicol}

\usepackage{enumitem} 					

\addto\extrasenglish{
	
} 

\usepackage[style=ieee,sorting=none,backend=biber]{biblatex}
\begin{document}
\thispagestyle{plain}
\pagestyle{plain}
\pagenumbering{arabic}
\begin{center}
	
\begin{spacing}{1}
{\LARGE \textbf{Proposal for a Comprehensive (Crypto) Asset Taxonomy}}
\end{spacing}
\bigskip
Thomas Ankenbrand \& Denis Bieri, Lucerne University of Applied Sciences and Arts;\\
Roland Cortivo \& Johannes Hoehener, Swisscom AG;\\
Thomas Hardjono, MIT Connection Science
\bigskip
\noindent
\\
\bigskip

\today

\end{center}

\begin{abstract}
Developments in the distributed ledger technology have led to new types of assets with a broad range of purposes. Although some classification frameworks for common instruments from traditional finance and some for these new, so-called cryptographic assets already exist and are used, a holistic approach to integrate both worlds is missing. The present paper\footnote{Our research is part of the FinTech programme supported by Finnova, Inventx, Swiss Bankers Prepaid Services, and Swisscom.} fills this research gap by identifying 14 attributes, each of which is assigned different characteristics, that can be used to classify all types of assets in a structured manner. Our proposed taxonomy, which is an extension of existing classification frameworks, summarises these findings in a morphological box and is tested for practicability by classifying exemplary assets like cash and bitcoin. The final classification framework can help to ensure that the various stakeholders, such as investors or supervisors, have a consistent view of the different types of assets, and in particular of their characteristics, and also helps to establish standardised terminology.
\end{abstract}

\section{Introduction}

Since the inception of the Bitcoin network in the year
2009, the space for cryptographic assets has developed
rapidly. The continuing technological innovation
in the underlying distributed ledger technology could
consequently lead to an increasing transformation of
traditional financial markets into crypto-based markets.
Although different asset classification frameworks
exist for both worlds, a holistic approach merging
both traditional finance and the crypto economy is
still lacking. This poses a challenge to the various
stakeholders such as investors or regulators in retaining
an overview of existing assets of different types
and, in particular, of their design and individual characteristics. In order to fill this lack of research, we propose a taxonomy for the systematic classification of all types of assets, be it of physical,
digital or tokenised nature.

\section{Literature Review}\label{chap:Literature Review}

The characteristics and properties of the most common
types of financial instruments such as stocks, bonds,
and derivatives have been the subject of research for
some time, not only in the academia, but also in the
industry. Therefore, a wide range of publications exist
that deal with the functioning of these different instruments
in a structured way.
\newline

One framework defining the structure and format for the classification of financial instruments (CFI) was first proposed by the International Organization for Standardization (ISO) in the year 1997. The last revised version of the framework is called ISO 10962:2019 and was published by ISO in 2019. It seeks to provide a standard for identifying the type of financial instrument and its main high-level features in the form of specific codes consisting of six alphabetical characters, and should thus help to standardise country and institution-specific terminology in relation to financial instruments \cite{iso2019}. The first character of the CFI code indicates the main category of financial instruments. These include equities, collective investment vehicles, debt instruments, different types of derivatives, and others.\footnote{For a detailed description of each category, see \cite{iso2019}.} The second character of the CFI code indicates multiple subclasses in a given main category, called groups. Equities, for example, are divided into the groups common/ordinary shares, preferred/preference shares, and common/ordinary convertible shares, among other groups. The last four characters of the CFI code define the specific attributes of a financial instrument and depend on the group to which the asset is allocated. For financial instruments in the group common/ordinary shares from the "equities" main category, relevant attributes include voting rights, ownership, payment status, and form. These attributes come with predefined possible values that determine the final code of a financial instrument \cite{iso2019}. For other groups such as bonds from the "debt instruments" category, alternative attributes, e.g., the type of interest or guarantee, are of relevance.
\newline

A second framework for classifying financial instruments is proposed by Brammertz and Mendelowitz \cite{brammertz2018digital}.
Their so-called ACTUS taxonomy is based on the
specific nature of financial contracts and in particular
on their cash flow profiles and seeks to create a global
standard for the consistent representation of financial
instruments. It distinguishes between financial contracts,
which in turn are split into the subcategories of
basic contracts and combined/derivatives contracts on
the one hand, and credit enhancement on the other.
Basic contracts consist of fixed income and index-based
products, whereas combined/derivative contracts comprise
symmetric financial products, options, and securitisation
products. The second main category of the
ACTUS taxonomy, i.e., credit enhancement, includes
guarantee contracts, collateral contracts, margining
contracts, and repurchase agreements. The standard
is implemented on the SolitX platform with a technical
API layer and DLT adapter for transaction systems and
accounting, and in the AnalytX architecture for risk
management analysis, simulations, asset and liability
management, and business planning \cite{Swisscom2019}.
\newline

The standards proposed by \cite{iso2019} and \cite{brammertz2018digital} show that sophisticated classification frameworks for traditional financial assets exist, which are used in practice. For cryptographic assets, on the contrary, the characteristics
of many tokens in various respects, for example
in terms of regulation, utility or valuation, were
and are still largely ambiguous and hard to measure.
Several initiatives from governments, the academia,
and the industry have sought to reduce these uncertainties
by systematically structuring the hundreds of
existing tokens based on predefined criteria. 
\newline

The Swiss Financial Market Supervisory Authority (FINMA), for example, issued guidelines for enquiries regarding the
regulatory framework for initial coin offerings in early
2018, in which it distinguishes between three types of
tokens, i.e., payment tokens, utility tokens, and asset
tokens, based on the underlying economic purpose
\cite{FINMA2020}. Whether a particular token is a financial instrument and thus would be subject to certain laws and regulations depends on its economic function and the rights associated with it. Other jurisdictions, such as the European Union, Israel, Malta, and the United Kingdom, follow a similar classification approach, although their terminologies differ to some extent \cite{blandin2019global}. Additionally, some jurisdictions follow the approach that the three main types of tokens are not necessarily mutually exclusive. Rather, there are also hybrid forms that share characteristics of two or three main types. Accordingly, particular cryptographic assets could thus, for example, have certain characteristics of both payment and utility tokens. 
\newline

In April 2019, the U.S. Securities and Exchange Commission (SEC) through its strategic hub for financial innovation, FinHub, published guidelines to determine whether a digital asset, which may be a cryptographic asset, is an investment contract, i.e. an agreement whereby one party invests money in a common enterprise with the expectation of receiving a return on investment. This assessment is done by applying the so-called Howey test. If an investment agreement exists, the digital asset is classified as a security and therefore U.S. federal securities laws apply and must be considered by issuers and other parties involved in, for example, the marketing, offering, sale, resale or distribution of the respective asset \cite{sec2019}. Other jurisdictions, e.g., Ireland, follow a similar approach of classifying cryptographic assets based on their qualification as a security \cite{lawlibrary2019}. However, the Howey Test is to be understood less as a classification framework but more as a decision aid as to whether a cryptographic asset represents a security or not.
\newline

An academically based classification framework for cryptographic assets, which goes beyond the legal perspective and also takes technological and economic aspects, among others, into account was carried out by Oliveira et al. \cite{oliveira2018token}. By applying 
a design science research approach, including 16 interviews
with representatives of projects with blockchain-based token systems, the paper derives a
token classification framework for cryptographic assets
that can be used as a tool for better informed decision
making when using tokens in blockchain applications.
Their final classification framework consists of the 13 attributes class, function, role,
representation, supply, incentive system, transactions,
ownership, burnability, expirability, fungibility, layer,
and chain, each of which include a set of defined characteristics. 
\newline

A similar framework was developed by Ballandies
et al. \cite{Ballandies2018}. The authors established a classification
framework for distributed ledger systems consisting
of a total of 19 descriptive and quantitative
attributes with four dimensions (distributed ledger, token,
action, and type). The attributes comprise the distributed
ledger type, origin, address traceability, Turing
completeness, and storage in the distributed ledger dimension,
underlying, unconditional creation, conditional
creation, transferability, burn, and supply in the
token dimension, action fee, read permission, and actor
permission in the action dimension, and fee, validate
permission, write permission, proof, and type in
the consensus dimension. The framework was derived
from feedback from the blockchain community.
\newline

Three further classification frameworks for cryptographic
assets that were strongly driven by the industry
are those proposed by the consulting firm MME, the International Token
Standardization Association (ITSA), and the Ethereum
Enterprise Alliance (EEA). 
\newline

The framework by MME was
published in May 2018 and focuses on the legal properties
and risk assessment of cryptographic assets. The paper’s
resulting classification is based on a token’s function
or main use, alongside other criteria such as the existence
of a counterparty, as well as its type and/or the
underlying asset or value. The final archetypes of cryptographic
assets are native utility tokens, counterparty
tokens, and ownership tokens, which are each subject to
additional subcategories of token types \cite{mueller2018conceptual}. 
\newline

The International Token Classification (ITC)
framework by the ITSA comprises an economic, technological,
legal, and regulatory vertical each containing a
set of subdimensions with different attributes. The economic
and technological verticals include three subdimensions
each, which refer to a token’s economic purpose,
its target industry, and the way of distribution, and
the technological setup, consensus mechanism, and
technological functionality, respectively. The legal vertical
includes the two subdimensions legal claim and issuer
type, whereas the regulatory vertical focuses on assessing
a tokens regulatory status in the US, China, Germany,
and Switzerland. Over all verticals, a total of twelve
subdimensions are defined, though ITSA plans to define
further subdimensions in the future. Concerning the evaluation
of these individual subdimensions, as of September
2019, the ITC framework already provided detailed
information on four of the twelve subdimensions, namely
for the economic purpose, industry, technological setup,
and legal claim. The classification into these four subdimensions
was compiled in a database covering more
than 800 cryptographic tokens. Besides the classification
framework and the corresponding database, the ITSA
also introduced a nine digit unambiguous identifier for
each token, the so-called International Token Identification
Number, short ITIN \cite{ITSA}. 
\newline

The third industry-driven framework for classifying cryptographic tokens
was published by the EEA in November 2019. Their proposed
Token Taxonomy Initiative (TTI) distinguishes between
five characteristics a token can possess. The first
characteristic is the token type and refers to whether a
token is fungible or non-fungible. The second characteristic,
the token unit, distinguishes between the attribute
of being either fractional, whole or singleton and indicates
whether a token is subdivisible or not. The value
type, as the third characteristic, can assume the attribute
of being either of an intrinsic value, i.e., the token itself is
of value (e.g., bitcoin), or a reference value, i.e., the token
value is referenced elsewhere (e.g., tokenised real estate).
Characteristic four, the representation type, comprises
the attribute of being common or unique. Common tokens,
on the one hand, share a single set of properties,
are not distinct from one another and are recorded in a
central place. Unique tokens, on the other hand, have
unique properties and their own identity, and can be
traced individually. The fifth and last characteristic is the
template type and classifies tokens as either single or
hybrid and refers to any parent/child relationship or dependencies
between tokens. Unlike single tokens, hybrid
tokens combine parent and child tokens in order to
model different use cases. In addition, the TTI provides
measures in order to promote interoperability standards
between different blockchain implementations
\cite{EEAEEA2019}.

\section{The (Crypto) Asset Taxonomy}\label{chap:The (Crypto) Asset Taxonomy}

Building on the literature review in Chapter \ref{chap:Literature Review}, this chapter proposes a holistic framework for the classification of assets.
Unlike existing classification frameworks, our asset
taxonomy aims to classify all existing types of assets,
i.e., assets from both traditional finance as well as the
crypto economy, based on their formal characteristics. Furthermore, the taxonomy introduces a terminology that is suitable for both traditional and the crypto assets. A morphological box is chosen as the methodological approach in order to be able to take
the multi-dimensionality of the matter into account.
The taxonomy is illustrated in Appendix~\ref{appendix:a}. In total, we identify 14 different attributes based on which all types of assets can be classified. They include claim structure, technology,
underlying, consensus/validation mechanism, legal
status, governance, information complexity, legal
structure, information interface, total supply, issuance, redemption, transferability, and fungibility, with each attribute comprising a set of at least two characteristics. Note that certain attributes in the frameworks discussed in Chapter \ref{chap:Literature Review} subsume some of the attributes presented here. Hence, our 14 attributes factorize these superordinate attributes to make them universally applicable. Table \ref{tab:table1} breaks down the 14 attributes in terms of their inclusion in the publications discussed in Chapter \ref{chap:Literature Review}. The first column shows the attribute labels of the taxonomy we propose. Column two to ten refer to the publications discussed, where an "x" indicates that the corresponding attribute is either explicitly or implicitly considered in the classification framework given in row one. Note that the terminology regarding a particular attribute differs across these publications, for example, because they focus on different types of assets. The terminology we propose generalises these terms to ensure compatibility across all types of assets, thus creating a common linguistic understanding. Also note that due to the extension of the taxonomy to traditional
assets, some DLT-specific attributes/characteristics in the publications discussed are summarised or generalised,
while new attributes/characteristics were added in order to enable the mapping of traditional asset types. Overall, Table \ref{tab:table1} shows that each of the existing frameworks covers certain attributes determined by the specific focus or objective of the publication. The framework of FINMA \cite{FINMA2020}, for example, focuses on regulatory aspects, and thus predominantly includes corresponding attributes, i.e., claim structure, legal status, and legal structure. Other frameworks, for example the one published by the EEA \cite{EEAEEA2019}, focus more on technological aspects or the design of token features. Overall, none of the frameworks discussed in Chapter \ref{chap:Literature Review} covers the full range of formal attributes identified in our taxonomy. However, our taxonomy is generally confirmed by the existing literature, as each attribute is considered in at least one of the existing classification frameworks. The degree of agreement with the classification framework we propose varies, however. While the publication of ISO \cite{iso2019} covers four attributes of our taxonomy, the publications of Oliveira et al. \cite{oliveira2018token} and Ballandies et al. \cite{Ballandies2018} cover ten. There are also differences in coverage from an attribute perspective. While the underlying of an asset is of relevance in all frameworks analysed, the attributes information interface and fungibility are only covered by two. The taxonomy we propose therefore goes further than the existing classification frameworks, firstly because it is independent of the type of assets to be classified and secondly because it contains additional attributes and characteristics. Since some of these attributes and characteristics are not intuitively clear, they are explained in more detail in the following:\\

\begin{table*}
	\caption{Coverage of the 14 attributes in existing classification frameworks}
	\label{tab:table1}
	\centering
	\begin{tabularx}{\textwidth}{@{}l*{8}{C}c@{}}
		\toprule
		Attribute     & ISO \space \space \cite{iso2019} & B.\&M. \cite{brammertz2018digital} & FINMA \cite{FINMA2020}  & O. et al. \cite{oliveira2018token} & B. et al. \cite{Ballandies2018} & MME \cite{mueller2018conceptual} & ITSA \cite{ITSA} & EEA \cite{EEAEEA2019} \\ 
		\midrule
		Claim structure
		&    x   &   x    &    x    &  x   &    &   x  &    x  &  &     \\ 
		Technology
		&        &      &   x    &   x  &   x &   x  &  x    &   x     &  \\ 
		Underlying
		&    x    &   x   &  x      &  x  & x  &  x  &   x  &   x       &  \\ 
		Consensus/Validation mechanism
		&        &      &      &    & x &  x  &    x &          &  \\ 
		Legal status
		&        &      &  x      &x   &   &  x  &  x   &          & \\ 
		Governance
		&        &      &      & x   & x &   x  &     &          &  \\ 
		Information complexity
		&   x     &   x   &       &     &  x &   &      &         & \\ 
		Legal structure
		&   x     &     x &   x     &    &   &   x  &    &         &  \\ 
		Information interface
		&        &      &       &     & x   &     &         &  x   &  \\ 
		Total supply
		&        &  x    &       &  x   &  x &    &         &  x  &  \\ 
		Issuance
		&        &   x   &       &  x  & x &     &        &   x   &  \\ 
		Redemption
		&        &    x  &      &  x   & x  &   &           &    x  &  \\ 
		Transferability	
		&        &      &    x    & x   & x  &   x  &         &   x   &  \\ 
		Fungibility	
		&        &      &        & x    &    &           &   &  x   &   \\ 
		\bottomrule
	\end{tabularx}
\end{table*}

\textbf{Claim structure:} Does the asset represent a claim, i.e., a demand for something due or believed to be due \cite{MerriamWebster}? 
\begin{itemize}[label={--}]
\itemsep0em 
\item No claim(s): The asset does not represent any kind of
claim.
\item Flexible claim(s): The asset represents certain claims,
the possession or exercise of which can depend on
certain conditions (e.g., catastrophe bonds).
\item Fixed claim(s): The asset represents claims which can
neither be restricted nor restrained under any condition (e.g., fixed income).
\end{itemize}
\bigskip

\textbf{Technology:} Which technology is the asset based on?
\begin{itemize}[label={--}]
\itemsep0em 
\item Physical: The asset exists in a physical form (e.g., gold bullion).
\item Digital: The asset exits in a digital form, but is not
based on the distributed ledger technology (e.g., electronic share).
\item Distributed ledger technology: The asset is based on
the distributed ledger technology, structured either
as a native token, i.e., a token that is native to a specific
blockchain, or as a protocol token, i.e., a token
issued on an existing blockchain protocol \cite{oliveira2018token} such as,
for example, ERC-20 or ERC-721 tokens for the
Ethereum blockchain.
\end{itemize}
\bigskip

\textbf{Underlying:} Which underlying or collateral is the asset’s
value based on?
\begin{itemize}[label={--}]
\itemsep0em 
\item No underlying: The asset’s value is not a derivative
of an underlying asset (e.g., bitcoin).
\item Company: The asset’s value represents a stake in a
company (e.g., equity).
\item Bankable asset: The asset’s value represents a bankable
asset, i.e., an asset that can be deposited in a bank or custody account (e.g., fiat currencies).
\item Cryptographic asset: The asset’s value represents a
cryptographic asset, i.e., an asset based on the distributed ledger technology (e.g., derivative of a cryptographic asset).
\item Tangible asset: The asset is in a physical form \cite{Kenton2019} (e.g., real estate).
\item Contract: The asset’s value represents a contract (e.g., license agreement).
\end{itemize}
\bigskip

\textbf{Consensus-/Validation-mechanism:} How is the agreement
on the finality (e.g., property rights or ownership
transfer) of the asset reached?
\begin{itemize}[label={--}]
\item Instant finality: Consensus is final. Mechanisms that
typically, but not necessarily, belong to the deterministic
type are, for example, notary services or
qualified written form.
\item Probabilistic finality: Consensus is not final, but reached
with a certain level of confidence. Mechanisms that
typically, but not necessarily, belong to the probabilistic
type are, for example, proof-of-work or proof-of-stake.
\end{itemize}
\bigskip

\textbf{Legal status:} What is the regulatory framework governing
the asset?
\begin{itemize}[label={--}]
\itemsep0em 
\item Regulated: There are regulatory requirements for the issuance, redemption and governance of the asset.
\item Unregulated: There is no specific regulatory framework for the issuance, redemption and governance of the asset.
\end{itemize}
\bigskip

\textbf{Governance:} In which way is the asset governed?
\begin{itemize}[label={--}]
\itemsep0em 
\item Centralised: The asset is governed by an authoritative
party or consortium.
\item Decentralised: The asset is governed without centralised
control (e.g., certain types of cryptographic assets such as bitcoin).
\end{itemize}
\bigskip

\textbf{Information complexity:\footnote{Note that the characteristics of this attribute build on each other, i.e., each characteristic contains additional information compared to the previous one.}} What type of information
complexity is associated with the asset?
\begin{itemize}[label={--}]
\itemsep0em 
\item Value: The asset represents a specific value (e.g., currencies).
\item Contract: The asset encompasses conditional information in addition to its value (e.g., coupon bonds or DLT-based smart contracts\footnote{Note that such (smart) contracts, as in the case of bitcoin, are not necessarily based on a Turing-complete system.}).
\item Turing completeness: The asset is based on a Turing-complete («universally programmable») computational model (e.g., Ethereum).
\end{itemize}
\bigskip

\textbf{Legal structure:} What is the legal form of the asset?
\begin{itemize}[label={--}]
\itemsep0em 
\item No legal structure: There is no legal structure governing
the asset.
\item Foundation: The asset is governed by a foundation/
trust structure.
\item Note/bond: The asset is structured as a note or bond.
\item Share: The asset is structured as a share.
\item Other\footnote{The characteristic "Other" subsumes the broad range of alternative legal structures for reasons of simplicity and practicability.}: The asset has an alternative legal structure (e.g., central bank money).
\end{itemize}
\bigskip

\textbf{Information interface:} How does the asset receive and/or send relevant information?
\begin{itemize}[label={--}]
\itemsep0em 
\item No interface: The asset has no kind of information interface.
\item Qualitative: The asset manages relevant information
indirectly through an authorised instance (e.g.,
general assembly).
\item Quantitative: The asset manages relevant information
from authorised sources automatically (e.g., IoT
sources or oracle interfaces in the case of DLT-based smart contracts).
\end{itemize}
\bigskip

\textbf{Total supply:} To which limit can the asset be generated?
\begin{itemize}[label={--}]
\itemsep0em 
\item Fixed: The total supply of the asset is fixed.
\item Conditional: The total supply of the asset is dependent
on predefined conditions.
\item Flexible: The total supply of the asset is managed
flexibly by authorised parties.
\end{itemize}
\bigskip

\textbf{Issuance:} How is the asset generated?
\begin{itemize}[label={--}]
\itemsep0em 
\item Once: After an initial issuance, no additional units of
the asset are issued.
\item Conditional: Additional units of the asset are issued
once predefined conditions are met (e.g., newly issued cryptographic assets through mining).
\item Flexible: Additional units of the asset can be issued
flexibly by authorised parties (e.g., increase in share capital).
\end{itemize}
\bigskip

\textbf{Redemption:} How is the number of outstanding assets reduced?
\begin{itemize}[label={--}]
\itemsep0em 
\item No redemption: The number of outstanding assets cannot
be reduced.
\item Fixed: The reduction of the number of outstanding
assets follows a predefined protocol.
\item Conditional: The reduction of the number of outstanding
assets is initiated once predefined conditions
are met.
\item Flexible: The reduction of the number of outstanding
assets can be carried out flexibly by authorised parties (e.g., share buyback).
\end{itemize}
\bigskip

\textbf{Transferability:} Can the asset’s ownership be transferred
to another party?
\begin{itemize}[label={--}]
\itemsep0em 
\item Transferable: The asset’s ownership can be transferred
to another party.
\item Non-transferable: The asset’s ownership cannot be transferred to another party, for example, by sale or giveaway (e.g., some types of registered securities).
\end{itemize}
\bigskip

\textbf{Fungibility:} Can the asset be interchanged with another
asset of the same type?
\begin{itemize}[label={--}]
\itemsep0em 
\item Fungible: The asset is substitutable with another asset
of the same type.
\item Non-fungible: The asset is not substitutable with another
asset of the same type (e.g., artwork).
\end{itemize}

\section{Classification Examples}\label{chap:Classification}

This subchapter seeks to test the above-mentioned
taxonomy with selected examples. First, the taxonomy is used to compare cash to bitcoin, as both are intended means of payment\footnote{Bitcoin is often considered to be a store of value, but the original intention is to provide an alternative means of payment.}. This comparison is followed by the classification of Ether, a utility token, Crowdlitoken, an asset token, CryptoKitties, and a traditional share.

\subsection{Comparison between Cash and Bitcoin}
As both cash and bitcoin follow the purpose of a means of payment, both assets share certain similarities (see Appendix~\ref{appendix:b}). Neither cash, in the case of a fiat money system, nor bitcoin have a direct underlying asset. The value of the two assets is rather based on the public's trust in the issuer of the currency or in the underlying technological protocol, respectively. There is also no oracle interface, i.e. no specific source that interacts (e.g., directly provides information) with cash or bitcoin. Since both assets are designed as cash equivalents, their units are transferable from one party to another and individual units are interchangeable. Besides these commonalities there are some significant differences. While cash represents a certain value which depends on the denomination, bitcoin is of contractual type as it is transferred via smart contracts in the Bitcoin-script which is not Turing-complete. Bitcoin is furthermore not subject to any type of legal claim and has no legal structure. In contrast, cash is regulated as legal tender under national law. Since cash is of physical form, consensus on its state is given deterministically by the owner of the asset. Bitcoin, on the contrary, is a digital representation of value based on the distributed ledger technology. It is the native token of the Bitcoin blockchain, the consensus of which is based on the proof-of-work mechanism and thus finality of the system is not guaranteed but only probabilistic. This implies a decentralised governance of the asset, which is in contrast to the centralised governance of cash by central banks. Both assets also differ in terms of their total supply as well as in their ways to manage the number of outstanding units. While the maximum supply of bitcoin is fixed at 21 million units, there is no such restriction for cash. The issuance of additional units of bitcoin is conditional on the mining of new blocks and reducing the number of outstanding units is not possible\footnote{It is possible to send units of bitcoin, or other cryptographic assets, to an address without a known private key, so that these units are no longer accessible. However, this does not reduce the number of total units in the system.}. The issuance and redemption of cash, on the contrary, is handled flexibly by central banks.{\tiny }

\subsection{Ether}
Ether (see Appendix~\ref{appendix:c}), which is classed as a utility token,
is the native token of the Turing-complete Ethereum
platform which is governed by the Ethereum Foundation
located in the Crypto Valley. The token itself is unregulated.
Although multiple decentralised systems which
can act as a quantitative oracle interface for the platform
exist, there are no legal claims and no underlyings
associated with the token. Consensus on the Ethereum
platform is, at the time of writing, achieved based on the
proof-of-work mechanism, and therefore is of a probabilistic
nature. As a consequence, the governance of
the token is decentralised. Like with bitcoin, the issuance
of Ether tokens is conditional on the creation of
new blocks, i.e., when miners get awarded with newly
mined units, and the destruction of existing units is not
possible. However, currently the total supply of Ether is
not limited. All Ether tokens are transferable between
parties and are fungible.

\subsection{Crowdlitoken}
Crowdlitokens (see Appendix~\ref{appendix:d}) are classed as asset tokens
and are tokenised real estate bonds, regulated
under the existing law. They are issued on the Ethereum
Blockchain under the ERC-20 standard and represent a
contract including fixed claims (e.g., voting and interest
payment). The token value is derived from the fundamental
value of the issuing company, and only indirectly
by its real estate portfolio. Due to the underlying
distributed ledger technology, consensus on the state
of the tokens is not final but only probabilistic. Crowdlitokens
are structured as notes/bonds. They are governed
in a centralised manner through a qualitative
oracle interface since token holders are allowed to vote
on changes proposed by the management. They can be
issued and burnt (e.g., through token buybacks) flexibly
by the corresponding company, implying a flexible token
supply. The Crowdlitoken is both transferable and
fungible, whereby only persons who have successfully
completed the KYC/AML audits can subscribe to the
bonds and exercise all rights relating to them.

\subsection{CryptoKitties}
CryptoKitties (see Appendix~\ref{appendix:e}), as the last example
from the crypto space, are collectible digital representations
of cats created on the Ethereum blockchain.
The corresponding smart contracts can generate
over four billion variations of phenotypes and genotypes
(CryptoKitties, 2019). CryptoKitties neither represent
claims against a counterparty, nor a specific
underlying. They are non-fungible - every cat is unique
- but transferable ERC-721 tokens, without any regulatory
or legal governance. Although the front-end as
a traditional web app is managed by the development
team, the token’s governance, e.g., ownership, is
decentralised. Since consensus of the underlying
Ethereum protocol is reached via a proof-of-work
mechanism, the finality of the state of a CryptoKitties
token is probabilistic. Also, there is no oracle interface
related to CryptoKitties tokens. The creation of additional
units is done by breeding two CryptoKitties, resulting
in a new unique kitty, represented by a newly
issued unique token, while destroying a unit is not
possible. The corresponding smart contract allows for
a total limit of around four billion cats that can be bred,
implying a fixed total supply.

\subsection{Traditional Share}
Traditional shares (see Appendix~\ref{appendix:f}), as the one example
from traditional finance, are either physical or
digital in nature and represent a contract including
fixed claims (e.g., voting and/or profit participation)
against a counterparty, with its fundamental value
also representing the underlying of the asset. Shares,
as a legal form, are governed in a centralised manner
and are subject to the existing law (e.g., national corporate
law), with the general assembly of shareholders
being the supreme organ of a stock corporation, i.e.,
acting as a qualitative oracle interface. Consensus on
the state of a share is deterministically given by the
share registry. The creation of new shares as well as
the reduction in share capital, for example through
share buybacks, is left to the general assembly of the
corporation. As a consequence, the total supply of
traditional shares is flexible. Shares are typically
transferable, with exceptions such as restricted shares,
and fungible, i.e., substitutable with other shares of
the same company.
\section{Conclusion}\label{chap:Conclusion}

Various classification frameworks for traditional and cryptographic assets already exist and are applied in practice. However, a universal approach linking the two worlds has not yet been developed. In this paper we fill this research gap by proposing a taxonomy that extends existing classification frameworks. We identify 14 different attributes that are supported by the existing literature and by which each type of asset can be properly classified. These attributes include the claim structure, technology, underlying, consensus-/validation mechanism, legal status, governance, information complexity, legal structure, information interface, total supply, issuance, redemption, transferability, and fungibility. With the help of a morphological box, various possible characteristics that an asset can have are identified and assigned to these attributes. In this way, our taxonomy bridges the gap between physical, digital, and cryptographic assets, where sometimes the same asset can appear in all three forms, thus creating clear terminology. Thanks to the methodical approach, the individual attributes can be expanded or broken down at any level of detail without changing the overall framework. The classification of selected assets, such as cash and bitcoin, has also shown that the proposed taxonomy is applicable in practice. In a next step, the robustness and practical relevance of the taxonomy could be further tested, for example by interviewing experts in the field.

\onecolumn
\onecolumn
\appendices

\section{Asset taxonomy}
\label{appendix:a}
\begin{figure}[H]
	\centering
	\includegraphics[width=0.7\linewidth]{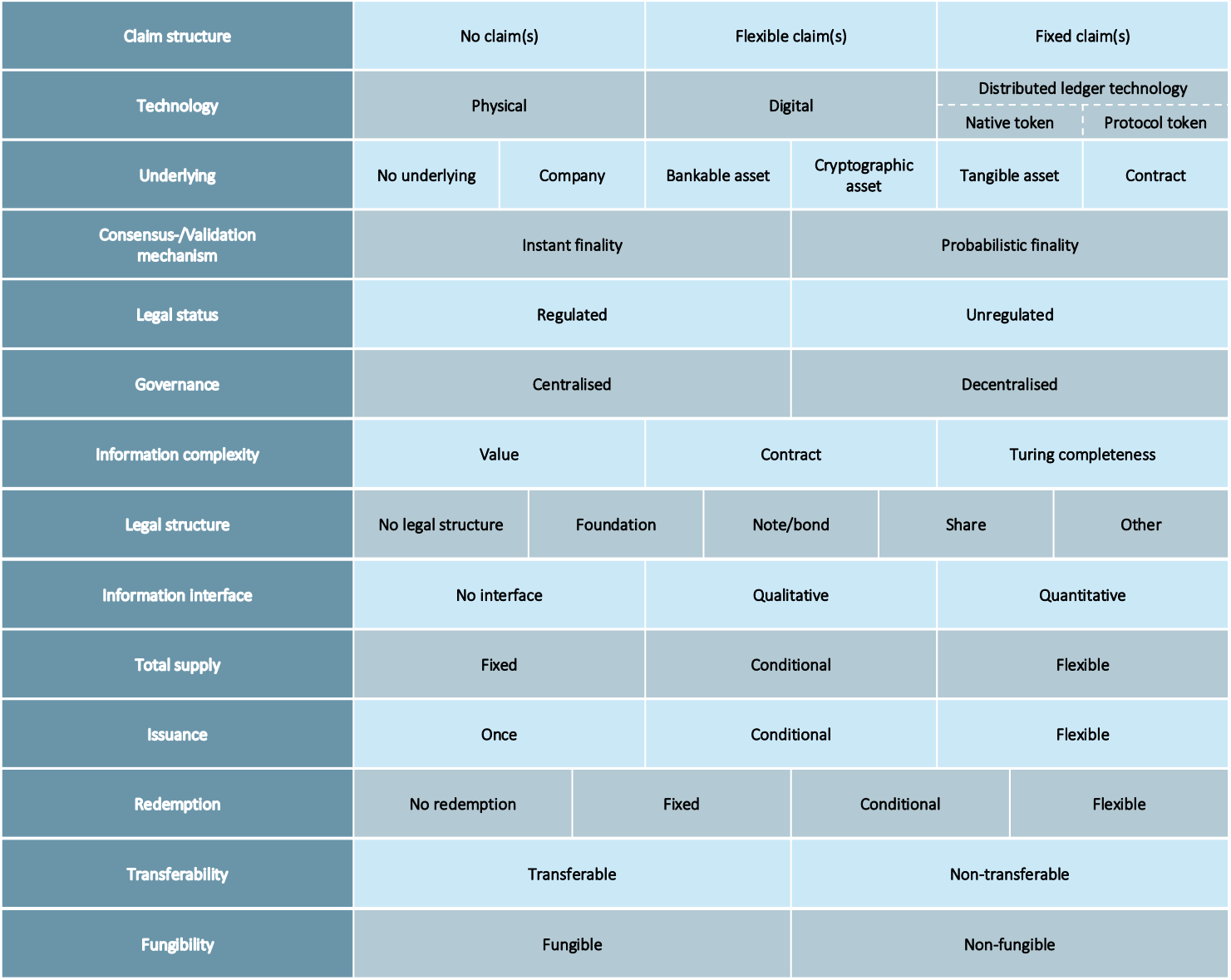}
\end{figure}

\section{Classification of cash (green) and bitcoin (orange)}
\label{appendix:b}
\begin{figure}[H]
	\centering
	\includegraphics[width=0.7\linewidth]{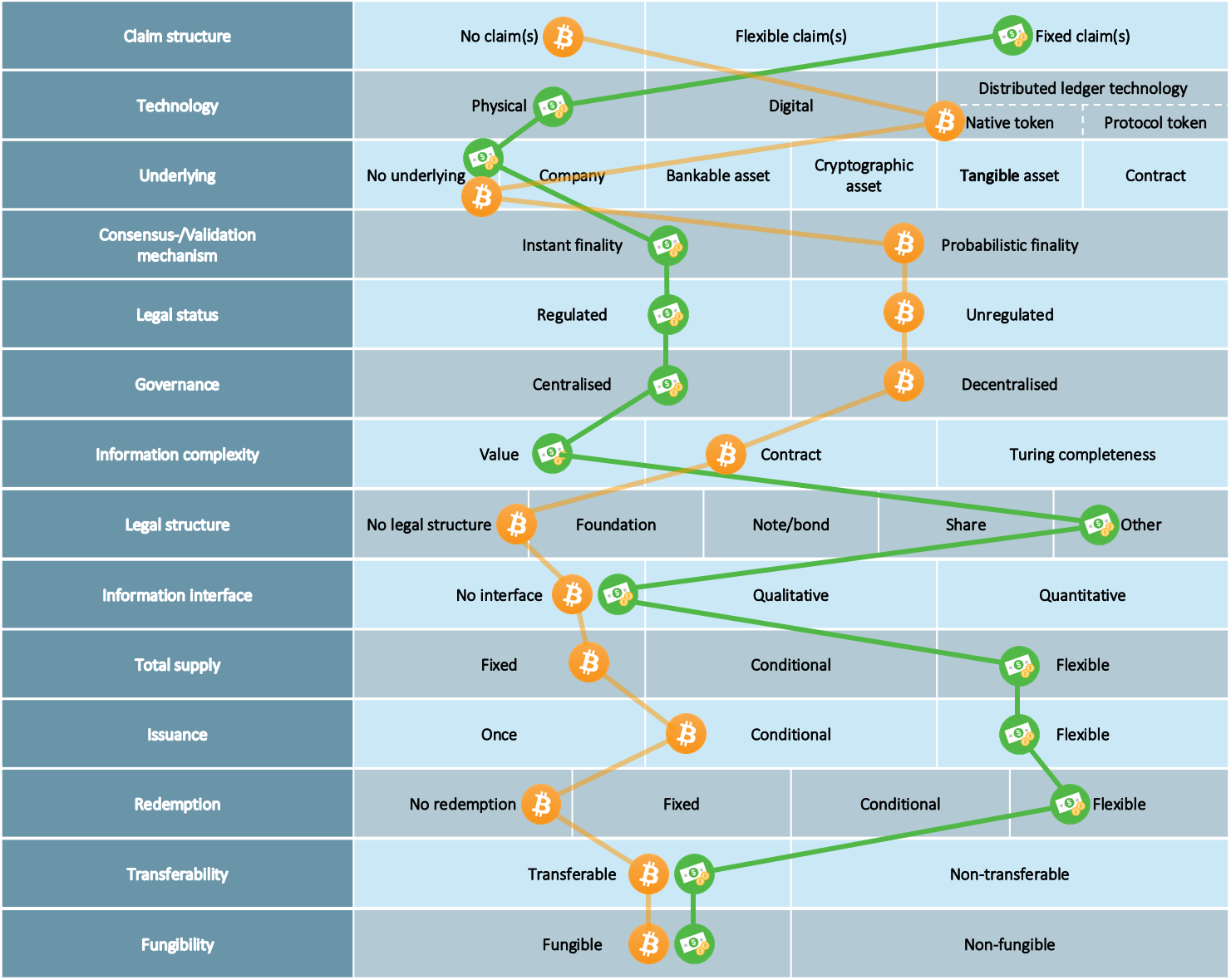}
\end{figure}

\section{Classification of Ether}
\label{appendix:c}
\begin{figure}[H]
	\centering
	\includegraphics[width=0.7\linewidth]{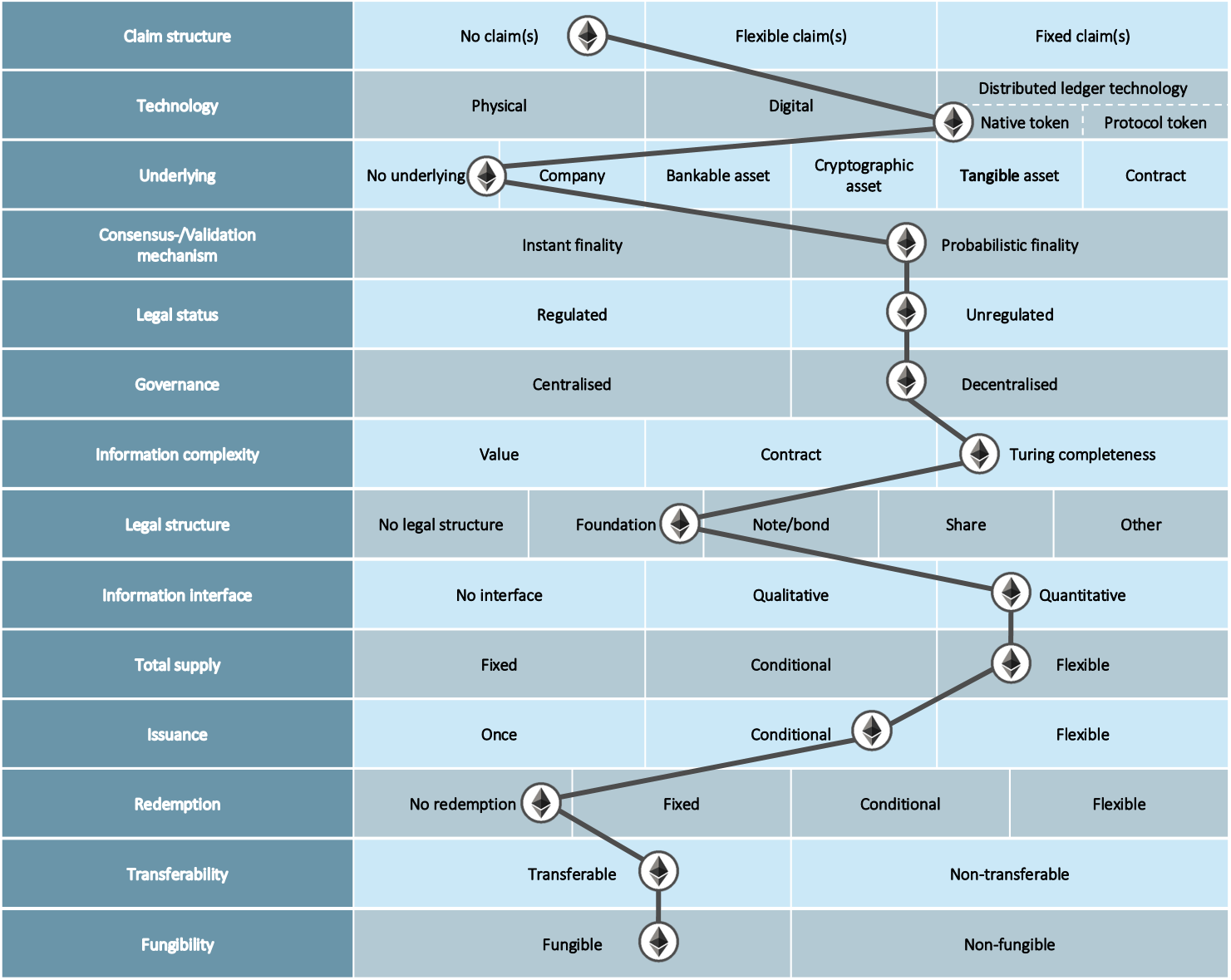}
\end{figure}

\section{Classification of a Crowdlitoken token}
\label{appendix:d}
\begin{figure}[H]
	\centering
	\includegraphics[width=0.7\linewidth]{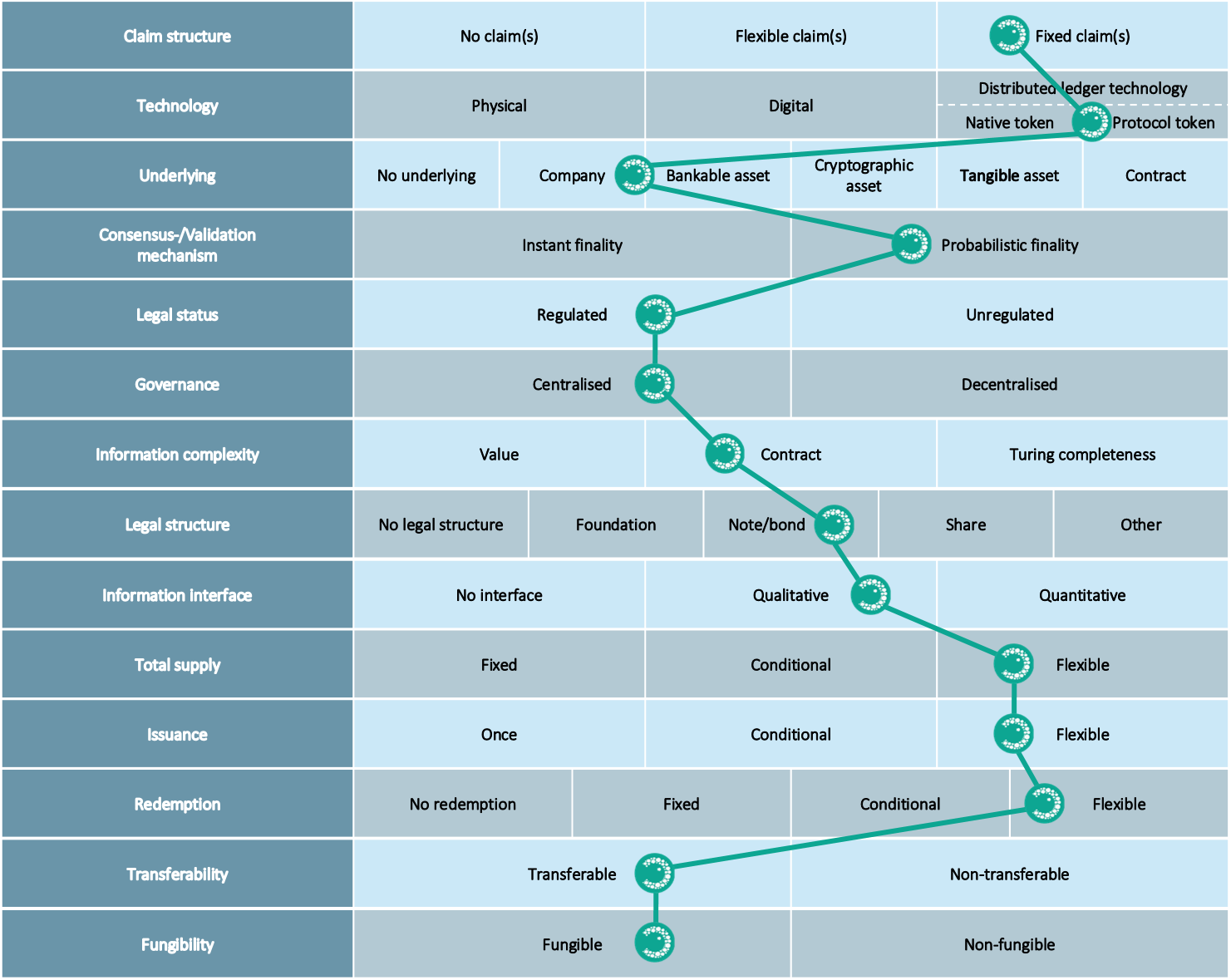}
\end{figure}

\section{Classification of a CryptoKitties token}
\label{appendix:e}
\begin{figure}[H]
	\centering
	\includegraphics[width=0.7\linewidth]{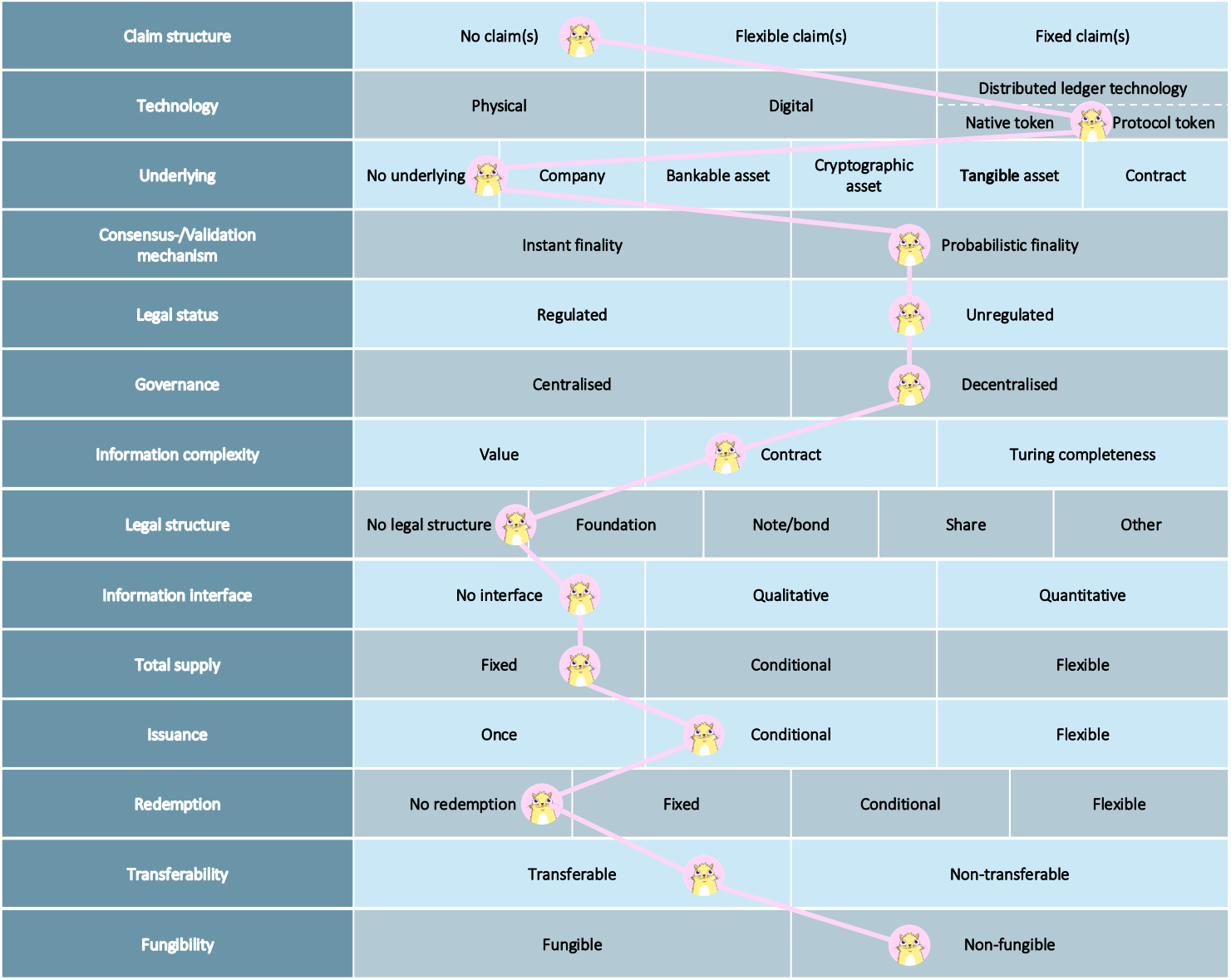}
\end{figure}

\section{Classification of a traditional share}
\label{appendix:f}
\begin{figure}[H]
	\centering
	\includegraphics[width=0.7\linewidth]{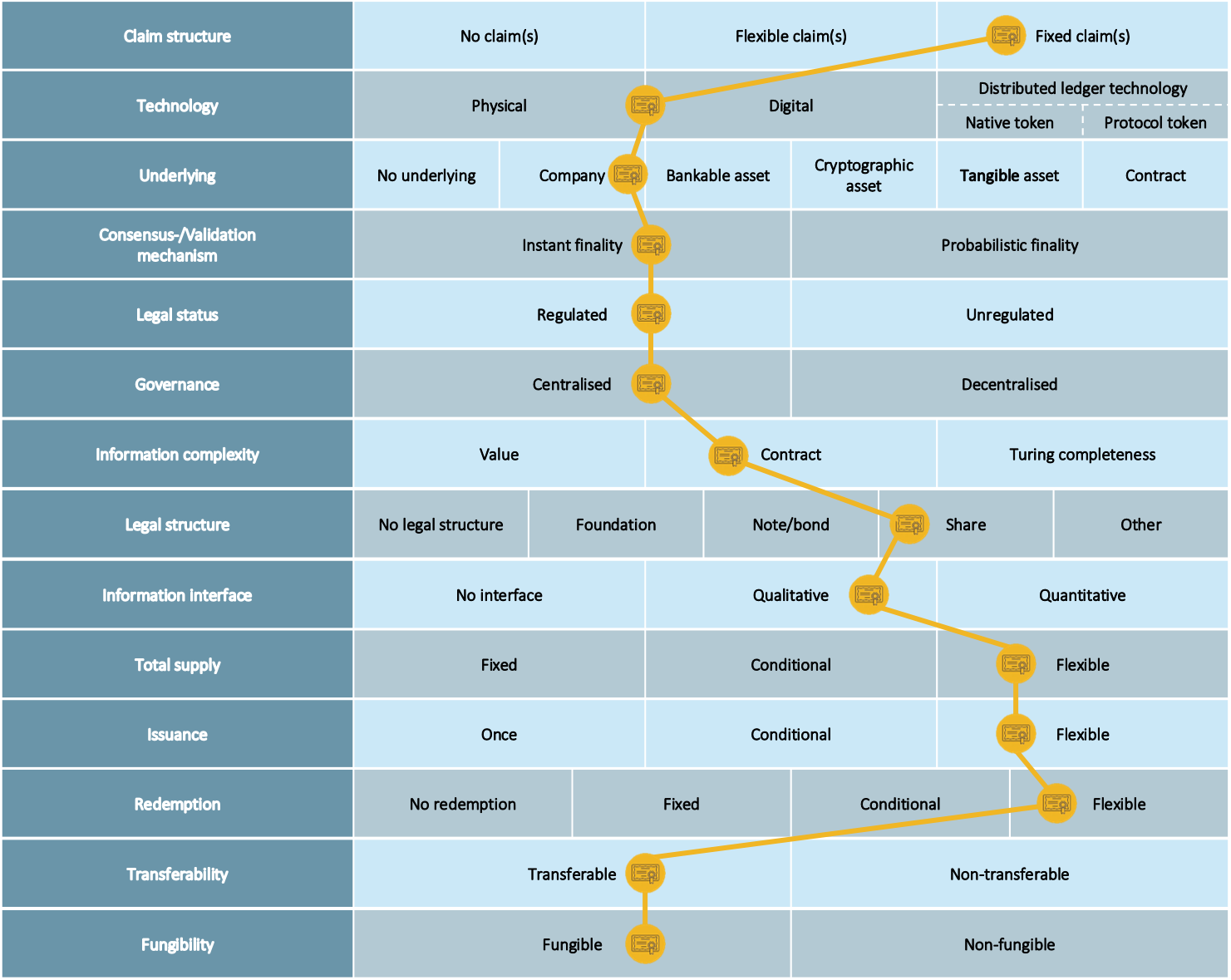}
\end{figure}

\twocolumn
\printbibliography

@Online{FINMA2020,
  author  = {{Swiss Financial Market Supervisory Authority FINMA}},
  date    = {2018},
  title   = {ICO Guidelines},
  url     = {https://www.finma.ch/en/~/media/finma/dokumente/dokumentencenter/myfinma/1bewilligung/fintech/wegleitung-ico.pdf?la=en},
  urldate = {2020-01-06},
}

@Article{Ballandies2018,
  author      = {Mark C. Ballandies and Marcus M. Dapp and Evangelos Pournaras},
  date        = {2018-10-30},
  title       = {Decrypting Distributed Ledger Design -- Taxonomy, Classification and Blockchain Community Evaluation},
  eprint      = {1811.03419v3},
  eprintclass = {cs.CY},
  eprinttype  = {arXiv},
}

@Online{ITSA,
  author  = {{International Token Standardization Association}},
  date    = {2019},
  title   = {Market standards for the global token economy},
  url     = {https://medium.com/@itsa_global/market-standards-for-the-global-token-economy-7d9f3d0cde37},
  urldate = {2020-03-25},
}

@Online{EEAEEA2019,
  author    = {{Enterprise Ethereum Alliance}},
  date      = {2019},
  title     = {Token Taxonomy Framework Overview},
  url       = {http://tokentaxonomy.org/wp-content/uploads/2019/11/TTF-Overview.pdf},
  urldate   = {2020-01-06},
}

@Online{sec2019,
  author  = {{The Law Library of Congress}},
  date    = {2018},
  title   = {Regulation of CryptocurrencyAround the World},
  url     = {https://www.loc.gov/law/help/cryptocurrency/cryptocurrency-world-survey.pdf},
  urldate = {2020-03-25},
}

@Online{iso2019,
  author    = {{International Organization for Standardization}},
  date      = {2019},
  title     = {Securities and related financial instruments — Classification of financial instruments (CFI) code, 10962:2019},
  url       = {https://www.sis.se/api/document/preview/80017455/},
  urldate   = {2020-03-25},
  publisher = {ISO, Geneva, Switzerland},
}

@Online{Kenton2019,
  author    = {Kenton, Will},
  date      = {2019},
  title     = {Tangible Asset},
  url       = {https://www.investopedia.com/terms/t/tangibleasset.asp},
  urldate   = {2020-03-25},
}

@Article{blandin2019global,
  author  = {Blandin, Apolline and Cloots, Ann Sofie and Hussain, Hatim and Rauchs, Michel and Saleuddin, Rasheed and Allen, Jason G and Cloud, Katherine and Zhang, Bryan Zheng},
  date    = {2019},
  title   = {Global cryptoasset regulatory landscape study},
  journal = {Available at SSRN},
}

@Online{mueller2018conceptual,
  author  = {Mueller, Lukas and Glarner, A and Linder, T and Meyer, SD and Furrer, A and Gschwend, C and Henschel, P},
  title   = {Conceptual Framework for Legal and Risk Assessment of Crypto Tokens},
  url     = {https://www.mme.ch/fileadmin/files/documents/180501_BCP_Framework_for_Assessment_of_Crypto_Tokens_-_Block_2.pdf},
  urldate = {2020-01-06},
  year    = {2018},
}

@Article{oliveira2018token,
  author    = {Oliveira, Luis and Zavolokina, Liudmila and Bauer, Ingrid and Schwabe, Gerhard},
  title     = {To token or not to token: Tools for understanding blockchain tokens},
  publisher = {University of Zurich},
  year      = {2018},
}

@Article{brammertz2018digital,
  author    = {Brammertz, Willi and Mendelowitz, Allan I},
  title     = {From digital currencies to digital finance: the case for a smart financial contract standard},
  journal   = {The Journal of Risk Finance},
  publisher = {Emerald Publishing Limited},
  year      = {2018},
}

@Online{lawlibrary2019,
  author  = {{The Law Library of Congress}},
  date    = {2019},
  title   = {Framework for “Investment Contract” Analysis of Digital Assets},
  url     = {https://www.sec.gov/corpfin/framework-investment-contract-analysis-digital-assets#_ednref2},
  urldate = {2020-03-26},
}

@Online{Swisscom2019,
  author  = {Swisscom},
  date    = {2019},
  title   = {SolitX: Smart Financial Contracts as a new approach to system support for banks},
  url     = {https://www.swisscom.ch/en/business/enterprise/themen/banking/solitx-smart-contracts.html},
  urldate = {2020-01-06},
}

@InCollection{MerriamWebster,
  author    = {{Merriam-Webster}},
  booktitle = {{Merriam-Webster online dictionary}},
  date      = {},
  title     = {claim},
  url       = {https://www.merriam-webster.com/dictionary/claim},
  urldate   = {26.05.2020},
}

\end{document}